\documentclass[12pt]{article}
\usepackage{latexsym}
\usepackage{graphicx}
\thicklines                         % thick straight lines for pictures (LaTeX)
\long \def \blockcomment #1\endcomment{}
\begin{document}           % End of preamble and beginning of text.

\baselineskip=0.33333in
%\begin{quote} \raggedleft TAUP 2846-2007
%\end{quote}
%\title{A Sample Document}  % Declares the document's title.
%\author{Leslie Lamport}    % Declares the author's name.
%\date{December 12, 1984}   % Deleting this command produces today's date.
\vglue 0.5in
%\maketitle                 % Produces the title.

\begin{center}{\bf Spin, Isospin and \\
Strong Interaction Dynamics}
\end{center}
\begin{center}E. Comay$^*$
\end{center}

\begin{center}
Charactell Ltd. \\
P.O. Box 39019, Tel Aviv 61390 \\
Israel
\end{center}
\vglue 0.5in
\vglue 0.5in
\noindent
%\vglue 0.2in
\vglue 0.5in
\noindent
Abstract:

The structure of spin and isospin is analyzed.
Although both spin and isospin are related to the same SU(2) group, they
represent different dynamical effects. The Wigner-Racah algebra is
used for providing a description of bound states of several Dirac
particles in general and of the proton state in particular.
Isospin states of the four $\Delta (1232)$ baryons are discussed.
The work explains the small contribution of quarks spin to the overall
proton spin (the proton spin crisis). It is also proved that the
addition of QCD's color is not required for a construction of an
antisymmetric state of the $\Delta ^{++} (1232)$ baryon.

%\vglue 0.5in
%\noindent

\newpage
\noindent
{\bf 1. Introduction}
\vglue 0.33333in

The isospin notion has been conceived by W. Heisenberg in 1932
(see [1], p. 106).
It aims to construct a mathematical basis that represents the
proton-neutron similarity with respect to
the strong nuclear force. Both spin and isospin have the same SU(2) group
structure. Thus, like spin multiplets of a quantum state, one combines
corresponding states of nuclear isobars in an isospin multiplet. For
example, the ground state of the $^{14}C$, $^{14}O$ and the $J^\pi=0^+$
excited state of $^{14}N$ are members of an isospin triplet.
Obviously, one must remember that isospin is a useful {\em approximation}
that neglects proton-neutron differences that are related to their mass and
their electromagnetic interactions.

Later developments have shown that the proton-neutron similarity
stems from the similarity between the {\em u, d} quarks. It follows that
the usefulness of isospin symmetry extends to particle physics.
For example, the three pions are members of an isospin triplet.
Due to historical development,
isospin notation takes different form in nuclear and particle physics.
Here $T$ and $I$ denote isospin in nuclear and particle physics,
respectively. In this work the symbol $T$ is used, mainly
because of the following
reason. In the case of spin, the symbols $J$ and $j$ denote total and
single particle angular momentum
operators, respectively. Similarly, the symbols $T$ and $t$ denote
the corresponding isospin operators. Thus, due to the same
underlying SU(2) group, isospin relations can be readily borrowed
from their corresponding spin counterparts. The operators $T$ and $t$ are
used in the discussion presented in this work.

This work examines states of electrons and quarks. These particles
have spin-1/2 and experimental data are consistent with their
elementary pointlike property. Evidently, a theoretical analysis of an
elementary pointlike particle is a much simpler task than that of
a composite particle. The discussion begins with an
examination of relevant properties of electronic states of atoms.
The mathematical structure of the
SU(2) group is used later for a corresponding analysis of isospin states.

Two important conclusions are derived from this analysis. First, it is well
known that quarks' spin carry only a small fraction of the entire proton's
spin [2]. This experimental evidence, which
is called the second EMC effect and
also the proton spin crisis, is shown here to be
an obvious result of the
multi-configuration structure of states of more than one Dirac particle.
Another result is that the anti-symmetric state of the $\Delta ^{++} (1232)$
baryon is well understood and there is no need to introduce a
new degree of freedom for its explanation. It means that the historical
starting point of the QCD construction has no theoretical basis.
(Below, the symbol $\Delta$ refers to this isospin quartet of baryons.)

Generally, in order to simplify notation, the specific value of
normalization factor is omitted from the expressions.
The second and the third sections analyze spin and isospin,
respectively. The fourth section provides an explanation for the proton spin
crisis. The fifth section explains the antisymmetric structure of
the $\Delta ^{++}$ baryon (without using color). The last section
contains concluding remarks.

\vglue 0.66666in
\noindent
{\bf 2. Spin States}
\vglue 0.33333in

A comprehensive discussion of angular momentum can be found in textbooks
[3]. In this short work some elements of this theory are mentioned
together with a brief explanation.
This is done for the purpose of arriving
rapidly at the main conclusions. A relativistic notation is used
and for this reason the $jj$ coupling [3] takes place.

Let us begin with a discussion of spin and spatial angular momentum.
These quantities are dimensionless and this property
indicates that they {\em may} be coupled. Now, the magnetic field
depends on space and time. Moreover, the theory must be consistent with
the experimental fact
where both spatial angular momentum and spin of an electron
have the same kind of magnetic field. Thus, it is {\em required} to
construct a relativistically
consistent coupling of these quantities. This is the theoretical basis
for the well known usage of spin and spatial angular momentum coupling in
the analysis of electronic states of atoms.

A motionless free electron is the simplest case and the spin-up
electron state is (see [4], p. 10)
\begin{equation}
\psi (x^\mu) = Ce^{-imt}
\left(
\begin{array}{c}
1 \\
0 \\
0 \\
0
\end{array}
\right),
\label{eq:FREEEL}
\end{equation}
where $m$ denotes the electron's mass.

A second example is the state of an electron bound to a
hypothetical pointlike very massive positive charge. Here the electron
is bound to a spherically symmetric charge $Ze$.
The general form of a $j^\pi$ hydrogen atom wave function is
(see [5], pp. 926, 927)
\begin{equation}
\psi (r\theta \phi) =
\left(
\begin{array}{c}
F {\cal Y}_{jlm} \\
G {\cal Y}_{jl'm}
\end{array}
\right),
\label{eq:HYDROGEN0}
\end{equation}
where ${\cal Y}_{jlm}$ denotes the ordinary $Y_{lm}$ coupled with a
spin-1/2 to $j$, $j=l\pm1/2$, $l'=l\pm 1$, $F,G$ are radial
functions and the parity is $(-1)^l$.

By the general laws of electrodynamics, the state must be an eigenfunction
of angular momentum and parity. Furthermore,
here we have a problem of {\em one}
electron (the source at the origin is treated as an inert object) and
indeed, its wave function $(\!\!~\ref{eq:HYDROGEN0})$ is an eigenfunction
of both angular momentum and parity
(see [5], p. 927).

The next problem is a set of $n$-electrons bound to an attractive
positive charge at the origin. (This is a kind of an ideal atom where
the source's volume and spin are ignored.) Obviously, the general laws
of electrodynamics hold and the system is represented by an
eigenfunction of the total angular momentum and parity $J^\pi$.
Here a single electron
is affected by a spherically symmetric attractive field {\em and} by the
repulsive fields of the other electrons. Hence, a single electron does
not move in a spherically symmetric field and it {\em cannot} be
represented by a well defined single particle
angular momentum and parity.

The general procedure used for solving this problem is to expand
the overall state as a sum of configurations. In every configuration,
the electrons' single particle angular momentum and parity
are well defined. These angular momenta are coupled
to the overall angular momentum $J$ and the
product of the single particle parity is the parity of the entire system.
The role of configurations has already been recognized in the early
decades of quantum physics [6]. An application of the first
generation of
electronic computers has provided a numerical proof of the vital
role of finding the correct configuration interaction required for a
description of even the simplest case of the ground state of the
two electron He atom [7]. The result has proved that several configurations
are required for a good description of this state and no configuration
dominates the others. This issue plays a very important role in
the interpretation of the state of the proton and of the $\Delta ^{++}$.

For example, let us write down the $0^+$ ground state He$_g$
of the Helium atom as a
sum of configurations:

\begin{eqnarray}
\psi (\mbox {He}_g) & = &
f_0(r_1)f_0(r_2) {\scriptstyle \frac {1}{2}^+ \frac {1}{2}^+} +
f_1(r_1)f_1(r_2) {\scriptstyle \frac {1}{2}^- \frac {1}{2}^-} +
f_2(r_1)f_2(r_2) {\scriptstyle \frac {3}{2}^- \frac {3}{2}^-} + \nonumber \\
 & &
f_3(r_1)f_3(r_2) {\scriptstyle \frac {3}{2}^+ \frac {3}{2}^+} +
f_4(r_1)f_4(r_2) {\scriptstyle \frac {5}{2}^+ \frac {5}{2}^+} + ...
\label{eq:HEATOM0}
\end{eqnarray}
Here and below, $f_i(r),\,g_i(r)$ and $h_i(r)$ denote the
two-component Dirac radial
wave function (multiplied be the corresponding coefficients).
In order to couple to $J=0$ the two single particle $j$ states must
be equal and in order to make an even total parity both must have
the same parity. These requirements make a severe restriction on
acceptable configurations needed for a description of the
ground state of the He atom.

Higher two-electron total angular momentum allows
a larger number of acceptable
configurations. For example, the $J^\pi = 1^-$ state of the He atom
can be written as follows:

\begin{eqnarray}
\psi (\mbox {He}_{1^-}) & = &
g_0(r_1)h_0(r_2) {\scriptstyle \frac {1}{2}^+ \frac {1}{2}^-} +
g_1(r_1)h_1(r_2) {\scriptstyle \frac {1}{2}^+ \frac {3}{2}^-} +
g_2(r_1)h_2(r_2) {\scriptstyle \frac {1}{2}^- \frac {3}{2}^+} + \nonumber \\
 & &
g_3(r_1)h_3(r_2) {\scriptstyle \frac {3}{2}^- \frac {3}{2}^+} +
g_4(r_1)h_4(r_2) {\scriptstyle \frac {3}{2}^- \frac {5}{2}^+} +
g_5(r_1)h_5(r_2) {\scriptstyle \frac {3}{2}^+ \frac {5}{2}^-} + \nonumber \\
 & &
g_6(r_1)h_6(r_2) {\scriptstyle \frac {5}{2}^+ \frac {5}{2}^-} ...
\label{eq:HEATOM1}
\end{eqnarray}

Using the same rules one can apply simple combinatorial calculations and
find a larger number of acceptable configurations
for a three or more electron atom. The main conclusion of this section
is that, unlike a quite common belief, there are only {\em three}
restrictions on configurations required for a good description of a $J^\pi$
state of more than one Dirac particles:
\begin{itemize}
\item[{1.}] Each configuration must have the total angular momentum $J$.
\item[{2.}] Each configuration must have the total parity $\pi$.
\item[{3.}] Following the Pauli exclusion principle,
each configuration should not contain two or more identical
single particle quantum states of the same Dirac particle.
\end{itemize}
These restrictions indicate that a state can be written as a sum of
many configurations, each of which has a well defined single particles
angular momentum and parity of its Dirac particles.

The mathematical basis of this procedure is as follows. Take the
Hilbert sub-space made of configurations that satisfy the three
requirements mentioned above and calculate the Hamiltonian matrix.
A diagonalization of this Hamiltonian yields eigenvalues and
eigenstates. These eigenvalues and eigenstates are related to
a set of physical states that have the given $J^\pi$.
As pointed out above, calculations
show that for a quite good approximation to a quantum state one needs a not
very small number of configurations and that no configuration has a
dominant weight. These conclusions will be used later in this work.

\vglue 0.66666in
\noindent
{\bf 3. Isopin States}
\vglue 0.33333in

Spin and isospin are based on the same mathematical group called SU(2).
Its three generators are denoted $j_x,j_y,j_z$. An equivalent basis is
(see [1], pp. 357-363)
\begin{equation}
j_+ = j_x + ij_y,\;\;
j_- = j_x - ij_y,\;\;j_z.
\label{eq:JZPM}
\end{equation}

All the $j$ operators mentioned above commute with the total $j^2$
operator. For this reason, if one of them operates on a member
of a $(2J+1)$ multiplet of an SU(2) irreducible representation then the
result belongs to this multiplet. The two $j_\pm$ operators are of
a particular importance. Thus, let $\psi_{J,M}$ denote a member of such
a multiplet and one finds
\begin{equation}
J_zJ_- \psi_{J,M} = (M - 1)J_-\psi_{J,M}.
\label{eq:JMINUS1}
\end{equation}
This relation means that $J_-$ casts $\psi_{J,M}$ into $\psi_{J,M-1}$
\begin{equation}
J_- \psi_{J,M} = \sqrt {J(J+1) - M(M-1)}\,\psi_{J,M-1},
\label{eq:JMINUS2}
\end{equation}
where the appropriate coefficient is written explicitly. Analogous
relations hold for the $J_+$ operator.

Let us turn to isospin. The required operators are
simply obtained by taking the mathematical
structure of spin and replacing the total spin operator $J$ and the
single particle spin operator $j$ by the corresponding isospin operators
$T,\,t$. (Here, like in the spin case, $M,\,m$ denote the eigenvalue of
$T_z,\,t_z$, respectively.)
The issue to be examined is
the structure of the isospin multiplet of the four baryons:
\begin{equation}
\Delta ^-,\, \Delta ^0,\, \Delta ^+,\,\Delta ^{++}.
\label{eq:4DELTAS}
\end{equation}
These $\Delta (1232)$ baryons have the lowest energy of the family
of the $\Delta$ baryons [8]. The $\Delta ^{++}$ baryon
has three $u$ quarks and $\psi _\Delta (uuu)$ denotes its state.
Therefore, its isospin state is $T=3/2,\,M=3/2$
and the isospin component of the wave function is symmetric with
respect to an exchange of any pair of quark.

Let us examine the operation of $T_-$ on $\Delta ^{++}$.
\begin{equation}
T_-\psi _\Delta (uuu) = (t_{1-} + t_{2-} + t_{3-})\psi _\Delta (uuu) =
\psi _\Delta (duu) + \psi _\Delta (udu) + \psi _\Delta (uud),
\label{eq:TMINUSDPP}
\end{equation}
where $t_{i-}$ operates on the ith quark. This is the way how one obtains
a yet unnormalized expression for the $\Delta ^+$ baryon from that of
$\Delta ^{++}$. A successive application of $T_-$ yields expressions
for every member of the isospin quartet $(\!\!~\ref{eq:4DELTAS})$.

Now, the $\Delta ^{++}$ state is symmetric with respect to its quark
constituents and the same property holds for the operator
$T_- = t_{1-} + t_{2-} + t_{3-}$. Hence, also the $\Delta ^{+}$
is symmetric with respect to its $uud$ quark states. This argument
proves that isospin space of {\em every} member of the baryonic quartet
$(\!\!~\ref{eq:4DELTAS})$ is symmetric. The same result can be obtained
from a different argument. Quarks are fermions and their overall
state must be antisymmetric with respect to an interchange of any
pair of quarks. Now, the isospin operators used above do not affect
other coordinates of quarks. It means that for every members of
the isospin quartet $(\!\!~\ref{eq:4DELTAS})$, the entire symmetry of
the other coordinates remain antisymmetric and the isospin coordinate
is symmetric.

The data confirms the similarity between members of an isospin
multiplet. Thus, for example, the mass difference between the
$\Delta ^0$ and $\Delta ^{++}$ baryons is less than 3 MeV
[8], whereas the mass difference between the $\Delta $
multiplet and the nucleons is about 300 MeV. This evidence
shows the goodness of the isospin notion, where strong interactions
dominate the state of members of an isospin multiplet and
the effect of all other interactions can be regarded as a small perturbation.

\vglue 0.66666in
\noindent
{\bf 4. The Proton Spin Crisis}
\vglue 0.33333in

The proton's $J^\pi = 1/2^+$
state is determined by three valence $uud$ quarks. The non-negligible
probability of the existence of an additional quark-antiquark pair
(see [1], p. 282) indicates that it is a highly relativistic system. The
discussion of section 2 holds for the spin-1/2 point-like quarks
and the expansion in configurations is a useful approach. Here the
three single particle $j^\pi$ represent the $uud$ quarks, in that
order. Evidently, each configuration must satisfy the three
requirement written few lines below $(\!\!~\ref{eq:HEATOM1})$.
However, the Pauli exclusion principle of restriction 3 does not
hold for the $d$ quark.
Thus, in analogy to $(\!\!~\ref{eq:HEATOM0})$ and
$(\!\!~\ref{eq:HEATOM1})$ one expands the
proton's wave function as a sum of terms of specific configurations.
A truncated expression for this expansion is shown below:
\begin{eqnarray}
\psi (uud) & = &
f_0(r_1)f_0(r_2)h_0(r_3) {\scriptstyle \frac {1}{2}^+ \frac {1}{2}^+ (0)
\frac {1}{2}^+} +
f_1(r_1)f_1(r_2)h_1(r_3) {\scriptstyle \frac {1}{2}^- \frac {1}{2}^- (0)
\frac {1}{2}^+} +                                               \nonumber \\
 & &
f_2(r_1)g_2(r_2)h_2(r_3) {\scriptstyle \frac {1}{2}^+ \frac {1}{2}^+ (1)
\frac {1}{2}^+} +
f_3(r_1)g_3(r_2)h_3(r_3) {\scriptstyle \frac {1}{2}^- \frac {1}{2}^- (1)
\frac {1}{2}^+} +                                               \nonumber \\
 & &
f_4(r_1)g_4(r_2)h_4(r_3) {\scriptstyle \frac {1}{2}^+ \frac {1}{2}^- (0)
\frac {1}{2}^-} +
f_5(r_1)g_5(r_2)h_5(r_3) {\scriptstyle \frac {1}{2}^+ \frac {1}{2}^- (1)
\frac {1}{2}^-} +                                                \nonumber \\
 & &
f_6(r_1)g_6(r_2)h_6(r_3) {\scriptstyle \frac {1}{2}^+ \frac {3}{2}^+ (1)
\frac {1}{2}^+} +
f_7(r_1)g_7(r_2)h_7(r_3) {\scriptstyle \frac {1}{2}^- \frac {3}{2}^+ (1)
\frac {1}{2}^-} +                                                \nonumber \\
 & &
f_8(r_1)g_8(r_2)h_8(r_3) {\scriptstyle \frac {1}{2}^+ \frac {1}{2}^+ (1)
\frac {3}{2}^+} +
f_9(r_1)g_9(r_2)h_9(r_3) {\scriptstyle \frac {1}{2}^- \frac {1}{2}^- (1)
\frac {3}{2}^+} +                                                \nonumber \\
 & &
f_a(r_1)g_a(r_2)h_a(r_3) {\scriptstyle \frac {1}{2}^- \frac {3}{2}^- (1)
\frac {1}{2}^+} +
f_b(r_1)g_b(r_2)h_b(r_3) {\scriptstyle \frac {1}{2}^+ \frac {3}{2}^- (1)
\frac {1}{2}^-} +                                                \nonumber \\
 & &
f_c(r_1)g_c(r_2)h_c(r_3) {\scriptstyle \frac {1}{2}^+ \frac {1}{2}^- (1)
\frac {3}{2}^-} + ...
\label{eq:PROTONWF}
\end{eqnarray}
The symbols ${\scriptstyle 0...9,a,b,c}$ are used for enumerating the terms.
Here, like in $(\!\!~\ref{eq:HEATOM0})$ and $(\!\!~\ref{eq:HEATOM1})$,
$f_i(r),\,g_i(r)$ and $h_i(r)$ denote the Dirac two-component radial
wave function of the $uud$ quarks, respectively
(multiplied be the corresponding coefficients). In each term,
the number in parentheses indicates how the two angular momenta of
the $uu$ quarks are coupled. Below, $J_{uu}$ denotes the value
of this quantity.

The following remarks explain the form of these terms.
An important issue is the coupling of the two $uu$ quark
that abide by the Pauli exclusion principle. For this reason, $J_{uu}$
is given explicitly in each term. Another
restriction stems from the rule of angular momentum addition. Thus,
for every term, the following relation must hold
in order to yield a total spin-1/2 for the proton: $J_{uu}=j_d \pm 1/2$.
These rules explain the specific structure of each term of
$(\!\!~\ref{eq:PROTONWF})$ which is described below.

In terms ${\scriptstyle 0,1}$
the two spin-1/2 are coupled antisymmetrically
to $J_{uu}=0$ and the two radial function are the same. In
terms ${\scriptstyle 2,3}$ these spins are coupled symmetrically to $J_{uu}=1$
and antisymmetry is obtained from the two orthogonal radial
functions. In terms ${\scriptstyle 4,5}$ the different
orbitals of the $uu$ quarks enable antisymmetrization. Thus, the two
spin-1/2 functions are coupled to $J_{uu}=0$ and $J_{uu}=1$,
respectively. The radial functions are not the same because of the
different orbitals. In terms ${\scriptstyle 6,7}$ the spins are
coupled to $J_{uu}=1$. In terms ${\scriptstyle 8,9}$
we have a symmetric angular momentum
coupling $J_{uu}=1$ and the antisymmetry is
obtained from the orthogonality of the radial function $f_i(r),\,g_i(r)$.
Terms ${\scriptstyle a,b}$ are analogous to terms ${\scriptstyle 6,7}$,
respectively. In term ${\scriptstyle c}$ the different $uu$ orbitals
enable antisymmetrization and they are coupled to $J_{uu}=1$.

A comparison of the expansion of the He atom ground state
$(\!\!~\ref{eq:HEATOM0})$ and that of the proton $(\!\!~\ref{eq:PROTONWF})$
shows the following points:
\begin{itemize}
\item[{1.}] If the expansion is truncated after the same value of a
single particle angular momentum then the number of terms in the proton's
expansion is significantly larger.
\item[{2.}] This conclusion is strengthened by the fact that the proton
has a non-negligible probability of an additional quark-antiquark
pair. An inclusion of this pair increases the number of acceptable
configurations.
\item[{3.}] Calculations show that the number of configurations required
for the ground state spin-0 of the two electron
He atom is not very small and that there is
no single configuration that dominates the state [7]. Now the proton is
a spin-1/2 relativistic particle made of three valence quarks.
Therefore, it is very
reasonable to assume that its wave function takes
a multiconfiguration form.
\end{itemize}

Using angular momentum algebra, one realizes that in most cases an
individual quark does not take the proton's spin direction. This
is seen on two levels. First, the upper and the lower parts of
the quark single particle function have
$l=j\pm 1/2$. Furthermore, the relativistic quark state indicates
that the coefficients of the
upper and the lower part of the Dirac four component function
take a similar size.
Hence, for the case where $j = l - 1/2$,
the Clebsch-Gordan coefficients [3] used for coupling the
spatial angular momentum and the spin indicate that
the spin of either the upper or the lower Dirac spinor
has no definite direction and that the coefficient of the spin down
is not smaller than that of the spin up (see [3], p. 519).

Let us turn to the coupling of the quark spins. The 3-quark
terms can be divided into two sets
having $j_{uu}=0$ and $j_{uu}>0$, respectively. For $j_{uu}=0$
one finds that the single particle $j_d=1/2$ and this spin is partially
parallel to the proton's spin. For cases where $j_{uu}>0$,
the proton's quark spins are coupled in a form where they take
both up and down direction so that they practically cancel each other.
The additional quark-antiquark pair increases spin direction mixture.
It can be concluded that the quark spin contribute a not very large portion
of the proton spin and the rest comes from the quark spatial motion.
This conclusion is supported by experiment [9].

\vglue 0.66666in
\noindent
{\bf 5. The State of the $\Delta ^{++}$ Baryon}
\vglue 0.33333in

In textbooks it is argued that without QCD, the state of the $\Delta ^{++}$
baryon demonstrates
a fiasco of the Fermi-Dirac statistics (see [10], p. 5). The argument
is based on the claim that the $\Delta ^{++}$ takes the lowest energy
state of the $\Delta $ baryons [11] and therefore, its spatial
wave function consists of three single particle symmetric s-waves of each
of its three $uuu$ quarks. Now the $J^\pi=3/2^+$ state of the $\Delta$
baryons shows that also their spin is symmetric. It means that the
$\Delta ^{++}$ is regarded to have space, spin and isospin symmetric
components of its wave function. As stated above,
textbooks claim that this outcome contradicts the Fermi-Dirac statistics.
However, using the physical issues discussed in this work and the following
energy level diagram of the nucleon and the $\Delta$ baryons,
it is proved that this textbook argument is incorrect.

\setlength{\unitlength}{1.0cm}
\begin{picture}(10,7)
\thicklines
\put(3,2.5){\line(1,0){1.5}}
\put(5,2.5){\line(1,0){1.5}}
\put(3.7,2.1){n}
\put(5.7,2.1){p}
\put(9.25,2.35){938}
\put(1,5.0){\line(1,0){1.5}}
\put(3,5.0){\line(1,0){1.5}}
\put(5,5.0){\line(1,0){1.5}}
\put(7,5.0){\line(1,0){1.5}}
\put(1.6,4.5){$\Delta ^-$}
\put(3.6,4.5){$\Delta ^0$}
\put(5.6,4.5){$\Delta ^+$}
\put(7.55,4.5){$\Delta ^{++}$}
\put(9.1,4.85){1232}
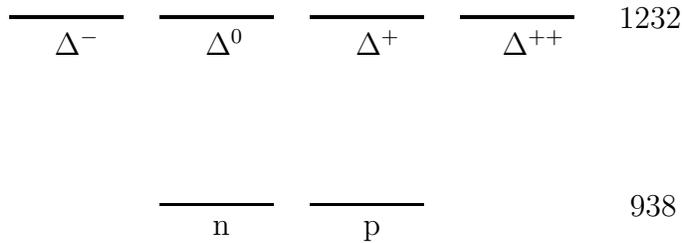
\put(1.0,1.25){{\em Fig. 1: Energy levels of the nucleon and the $\Delta$
isospin}}
\put(1.0,0.65){{\em multiplets (MeV).}}
\end{picture}

\begin{itemize}
\item As explained in section 3, all members of an isospin multiplet
have the same symmetry. Hence, if there is a problem with the
Fermi-Dirac statistics of the $\Delta ^{++}$ then the same problem
exists with $\Delta ^+$ and $\Delta ^0$. It follows that if the
above mentioned textbook argument
is correct then it is certainly incomplete.

\item The data described in fig. 1 shows that $\Delta ^+$ is an
excited state of the proton. Hence, its larger mass is completely
understood. Thus, there is no problem with the Fermi-Dirac statistics
of the $\Delta ^+$ baryon. Analogous relations hold for the neutron
and the $\Delta ^0$ baryons. Using the identical statistical state
of the four $\Delta$ baryons $(\!\!~\ref{eq:4DELTAS})$,
one realizes that there is no problem
with the Fermi-Dirac statistics of the
$\Delta ^{++}$ and the $\Delta ^-$ baryons.

\item The multi-configuration structure of a bound system of Dirac
particles is known for about 50 years [7]. In particular, the
multi-configurations structure
of all baryons (like in $(\!\!~\ref{eq:PROTONWF})$) proves that,
contrary to the above mentioned textbook argument (see [10], p. 5),
the single particle
spatial wave functions of the three $u$ quarks of the $\Delta ^{++}$
baryon {\em are not a pure s-wave}.

\end{itemize}

\vglue 0.66666in
\noindent
{\bf 6. Conclusions}
\vglue 0.33333in

This work uses the Wigner-Racah mathematical structure and proves two
very important points. It explains the small contribution of quark's spin
to the overall proton spin. Therefore, it eliminates
the basis for the proton spin crisis. It also proves that
everything is OK with the Fermi-Dirac statistics of the $\Delta ^{++}$
baryon. It follows that there is no need to introduce the QCD's
color degree of freedom in order to build an antisymmetric
wave function for this baryon.

%  ??????????????????????????????

\newpage
References:
\begin{itemize}
\item[{*}] Email: elicomay@post.tau.ac.il  \\
\hspace{0.5cm}
           Internet site: http://www.tau.ac.il/$\sim $elicomay

\item[{[1]}] D. H. Perkins {\em Introductions to High Energy Physics}
(Addison-Wesley, Menlo Park, 1987) 3rd edn.
\item[{[2]}] J Ashman et al. (EMC) Phys. Lett. {\bf B206}, 364 (1988).
\item[{[3]}] A. de-Shalit and I. Talmi, {\em Nuclear Shell Theory}
(Academic, New York, 1963).
\item[{[4]}] J. D. Bjorken
and S. D. Drell {\em Relativistic Quantum Mechanics} (McGraw, New York, 1964).
\item[{[5]}] A. Messiah, {\em Quantum Mechanics} (Dover, Mineola, 1999).
\item[{[6]}] H. A. Bethe, {\em Intermediate Quantum Mechanics} (Benjamin,
New York, 1964). (see p. 109).
\item[{[7]}] A. W. Weiss, Phys. Rev. 122, 1826 (1961)
\item[{[8]}] C. Amsler et al. (Particle Data Group),
Phys. Lett. {\bf B667}, 1 (2008).
\item[{[9]}] S.E. Kuhn, J.-P. Chen, E. Leader, Prog. Part. Nucl. Phys.
{\bf 63} 1 (2009).
\item[{[10]}] F. Halzen and A. D. Martin, {\em Quarks and Leptons}
(Wiley, New York, 1984).
\item[{[11]}] Today more than 10 different $\Delta$ baryonic multiplets
are identified [8].

\end{itemize}

\end{document}